\documentclass[fleqn,usenatbib]{mnras}

\usepackage{newtxtext,newtxmath}

\usepackage[T1]{fontenc}

\DeclareRobustCommand{\VAN}[3]{#2}
\let\VANthebibliography\thebibliography
\def\thebibliography{\DeclareRobustCommand{\VAN}[3]{##3}\VANthebibliography}

\usepackage{graphicx}	

\def \twomass {2MASS J01151085-7256102}
\def \scwdold {Swift~J004427.3-734801}
\def \scnew   {Swift~J004516.6$-$734703}
\def \scwd {Swift J011511.0-725611} 
\def \bexrb {BeXRB}
\def \bexrbs {BeXRBs}
\def \scubed {S-CUBED}
\def \swift {Swift}
\def \chandra {Chandra}
\def \xmm {XMM-Newton}

\title [Swift J011509.5-725610]{\scwd: Discovery of a rare Be Star / White Dwarf binary system in the SMC}

\author[J.~A. Kennea et al.]{J.~A. Kennea,$^{1}$\thanks{E-mail: jak51@psu.edu (JAK)}
M.~J. Coe,$^{2}$
P.~A. Evans$^{3}$,
L.~J. Townsend$^{4}$,
Z.~A. Campbell$^{1}$,
A. Udalski$^{5}$
\\
$^{1}$Department of Astronomy and Astrophysics, The Pennsylvania State University, 525 Davey Lab, University Park, PA 16802, USA\\
$^{2}$Physics \& Astronomy, The University of Southampton, SO17 1BJ, UK\\
$^{3}$University of Leicester, X-ray and Observational Astronomy Research Group, School of Physics \& Astronomy, University Road, Leicester LE1 7RH, UK\\
$^{4}$South African Astronomical Observatory, P.O Box 9, Observatory, 7935, Cape Town, South Africa\\
$^{5}$Astronomical Observatory, University of Warsaw, Al. Ujazdowskie 4, 00-478 Warszawa, Poland 
}
\date{Accepted 2021 September 9. Received 2021 August 31; in original form 2021 July 29}

\pubyear{2021}

\begin{document}
\label{firstpage}
\pagerange{\pageref{firstpage}--\pageref{lastpage}}
\maketitle

\begin{abstract}
We report on the discovery of \scwd, a rare Be X-ray binary system {(BeXRB)} with a White Dwarf {(WD)} compact object, in the Small Magellanic Cloud {(SMC)} by \scubed, a weekly X-ray/UV survey of the SMC by the Neil Gehrels Swift Observatory. Observations show an approximately 3 month outburst from \scwd, the first detected by \scubed{} since it began in 2016 June. \scwd{} shows super-soft X-ray emission, indicative of a White Dwarf compact object, which is further strengthened by the presence of an 0.871~keV edge, commonly attributed to O~\textsc{viii} K-edge in the WD atmosphere. Spectroscopy by SALT confirms the Be nature of the companion star, and long term light-curve by OGLE finds both the signature of a circumstellar disk in the system at outburst time, and the presence of a 17.4~day periodicity, likely the orbital period of the system. \scwd{} is suggested to be undergoing a Type-II outburst, similar to the previously reported SMC Be {White Dwarf binary (BeWD)}, \scwdold. It is likely that the rarity of known BeWD is in part due to the difficulty in detecting such outbursts due to both their rarity, and their relative faintness compared to outbursts in Neutron Star \bexrbs.
\end{abstract}




\section{Introduction}

Be/X-ray Binaries (\bexrbs) are high mass X-ray binaries (HMXB) consisting of typically a Neutron Star (NS) compact object and a main sequence star of B spectral type, showing strong emission line features in its optical spectrum, {most commonly hydrogen} Balmer lines,  {but also may show He and Fe emission lines \citep{Hanuschik1996}}. Although NS are the most common compact object type in these objects, in some rare cases the compact object in the system has been reported to be a Black Hole \citep{Munar-Adrover14}, or a White Dwarf (WD), most recently the WD containing \bexrb{} (hereafter BeWD) \scwdold{} discovered in the SMC \citep{coe2020}.  

\bexrbs{} are often discovered through two modes of X-ray emission, typically referred to as Type-I and Type-II outbursts. Type-I outbursts {($L_\mathrm{X} \approx 10^{36} - 10^{37}$~erg/s)}, the most common, occur when the compact object interacts with the Be star. This accretion occurs when the compact object passes through a circumstellar disc (CSD) which forms around the Be star. The presence of the CSD is strongly inferred from {the presence of excess infrared emission over that expected from the star}, which {brightens} when the CSD grows {\cite{Porter03}}. Type-I X-ray outbursts are transient, and their onset are linked to the growth of the CSD (e.g. \citealt{Kennea2020}). 

Type-II outbursts are typically at least an order of magnitude brighter than Type-I {($L_\mathrm{X} \gtrsim 10^{37}$~erg/s)}, sometimes reaching Super-Eddington luminosities (e.g. in SMC X-3, \citealt{Townsend2017}), and are not obviously linked to the orbital period of the source, with outbursts that can span multiple orbital periods \citep{Reig11}. The extended length Type-II outbursts suggest the formation of an accretion disc around the compact object, but the conditions required to form this are unclear, with suggestion that warping of the CSD plays a role (e.g. \citealt{Negueruela01}). 

When the compact object is a NS, an X-ray pulsar is typically seen, along with an X-ray spectrum which in the soft X-ray band ($0.5 - 10$~keV) is well described by a hard power-law spectrum with photon indices of $0-1$ typical. In the case of \bexrbs{} with WDs, the X-ray spectrum seen is more typical of a so-called Super Soft Source (SSS; e.g. \cite{Kahabka06}), in which the X-ray emission is thermal in nature and is predominantly emitting below 1~keV. 

The irregular dwarf galaxy the Small Magellanic Cloud (SMC) lies at a distance of 62.1 kpc \citep{graczyk2014}, with a very low X-ray absorption in the line-of-sight of $N_\mathrm{H} = 5.9 \times 10^{20}$~cm$^{-2}$ \citep{dl1990}. Although many X-ray binaries are known in the SMC, all are HMXB, and predominantly they are \bexrbs{}, which has made it a well studied area of the sky to look for \bexrbs{} in the past (e.g. \citealt{haberl2000, Haberl2016, CK2015}).

Motivated by the study of \bexrbs, but with more of a focus on long term transient behaviour, the \swift{} SMC Survey (\scubed; \citealt{kennea2018}) began in 2016. \scubed{} consists of a weekly survey of the optical extent of the SMC utilizing the Neil Gehrels Swift Observatory (\swift; \citealt{gehrels04}). The survey consists of 142 pointings with an exposure of 60s each. Data are taken utilizing \swift's X-ray Telescope (XRT; \citealt{burrows05}) in Photon Counting (PC) mode and Ultra-violet/Optical Telescope (UVOT; \citealt{Roming05}) utilizing the \textit{uvw1} filter. Due to the low background of XRT PC mode, 60s exposure allows for detection of an X-ray source in the SMC at a brightness of $1-2\%$ Eddington Luminosity for a $1.4~\mathrm{M}_\odot$ NS or brighter. 

{In this paper} we report the discovery, by \scubed{}, of a likely new \bexrb{} containing a WD compact object, \scwd. We present results of both \scubed{} and \swift{} target of opportunity (TOO) observations, as well as reporting on historical I-band light-curve from the Optical Gravitation Lensing Experiment (OGLE; \citealt{Udalski2003,Udalski2015}, and spectroscopic observations by the South African Large Telescope (SALT). 

\section{Observations}

\subsection{Discovery of \scwd}

\scwd{} was first discovered by the \scubed{} survey, and was reported by \cite{kennea2021}. The new source, internally designated SC1825 and subsequently named \scwd{} using the standard \swift{} naming scheme, was first detected in \scubed{} data taken on 2020 December 29, with an initial brightness of $0.05^{+0.07}_{-0.03}$ count~s$^{-1}$. No previous detection of this source was found in any of the \scubed{} observations going back to when the survey started on 2016 June 8. Subsequent observations saw the brightness of the source rise to $0.15^{+0.07}_{-0.03}$ count~s$^{-1}$ on 2021 January 5 \scubed{} observation, and then $0.19^{+0.08}_{-0.03}$ count~s$^{-1}$ in the 2021 Jan 12 observation. At this point, a \swift{} Target of Opportunity (TOO) request was triggered to obtain more detailed observations.

The most accurate X-ray position for \scwd, determined using the method described by \cite{Goad07}, was found to be R.A.(J2000) = $01^h 15^m 10.93^s$, Dec.(J2000) = $-72^\circ 56' 10.4''$ with an estimated uncertainty of $2.0''$ radius (90\% confidence). In UVOT data we find a single bright star inside the XRT error circle. A catalog search of this source revealed a known star, \twomass, with reported spectral types of O9IIIe \citep{gm2016} and O8-O9Ve \citep{lamb2013}, suggesting that this source was a new \bexrb{}. 

\subsection{\swift{} Observations}

\begin{figure}
	\includegraphics[width=9cm,angle=-0]{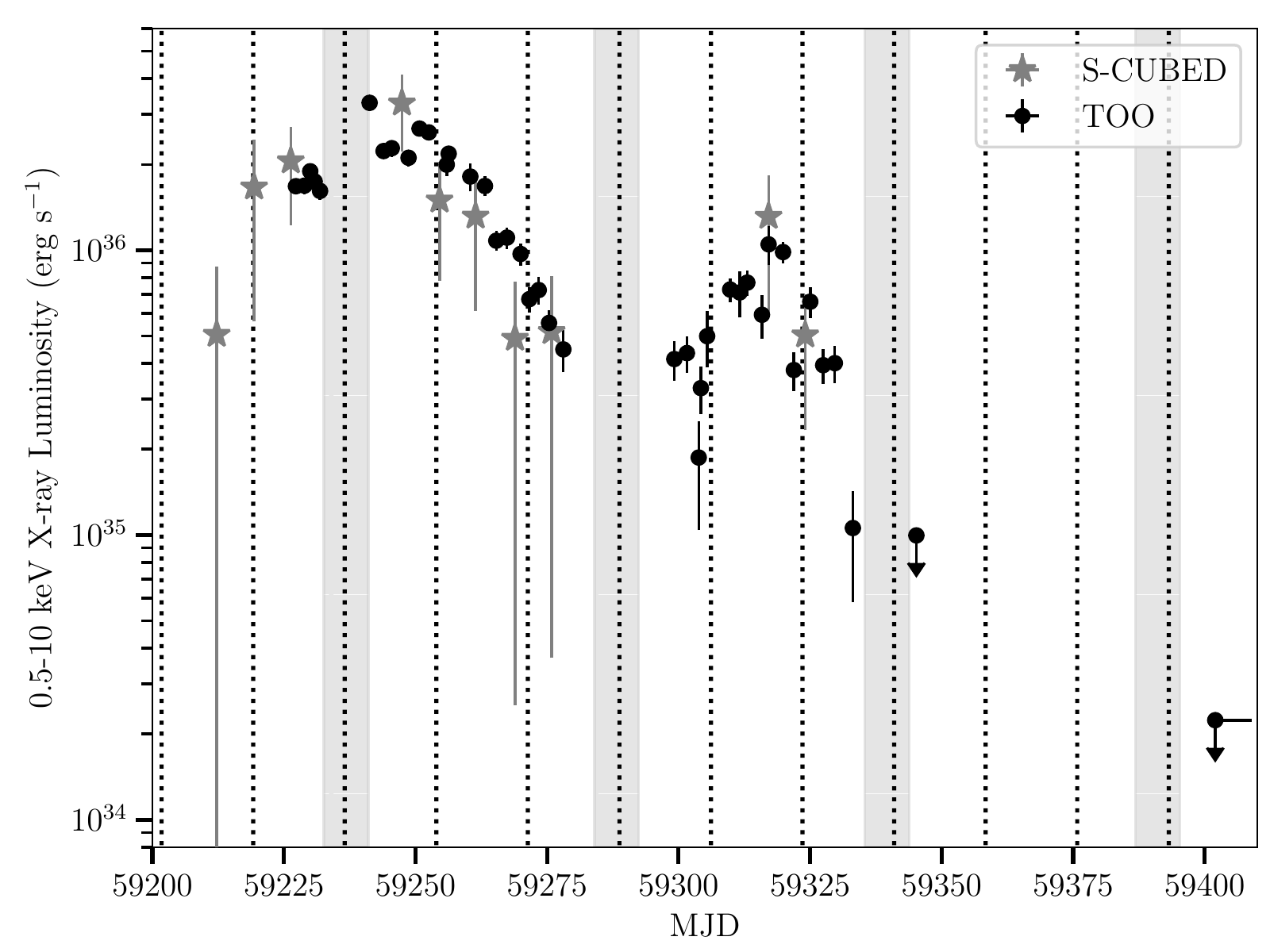}
    \caption{Combined \scubed{} and TOO X-ray light-curve of \scwd, showing the outburst period. Grey regions show periods of time when the source was not observable by \swift{} due to observing constraints. Dotted lines show the optical period found from OGLE data, using Eqn.~\ref{eq:1}.}
    \label{fig:xlc}
\end{figure}

The field containing \scwd{} has been regularly observed as part of the weekly \scubed{} survey \citep{kennea2018} starting on 2016 June 8th. As of 2021 July 27, it has been observed with \scubed{} on 199 occasions, with a median exposure time of 60s per observation. Although \scubed{} observations are planned weekly, observation gaps occur frequently due to \swift{} visibility or scheduling constraints. In addition to the \scubed{} observations, detection of the outburst lead to TOO observations of \scwd approximately every 1-2 days, starting on 2021 January 13 with the final exposure here taken on 2021 May 11. In addition, in 2021 July, late time observations totaling 5.08~ks exposure were taken post-outburst in order to determine a strong upper limit on quiescent emission from \scwd. A detailed list of TOO observations is given in Table~\ref{tab:x1} and the measurements are shown in Fig.~\ref{fig:xlc}. For all XRT observations the data were collected in PC mode. UVOT data were collected in \textit{uvw1} for \scubed{} data, and in the all three UV filters for TOO observations. In addition some of the TOO observations also collected data in $u$, $b$ and $v$ filters. Although the TOO observations aimed for twice daily cadence, due to scheduling and spacecraft constraints, there are gaps in the coverage, especially during periods of time when the source was near the \swift{} orbit pole, when observations are not possible for up to 10 days. 

\begin{table}
\begin{tabular}{rllr}
\hline
       ObsID & Begin (UTC)         & End (UTC)           &   Exp. (s) \\
\hline
 00013984001 & 2021-01-13 05:02:02 & 2021-01-13 08:30:54 &       4440 \\
 00013984002 & 2021-01-14 20:47:02 & 2021-01-14 22:40:06 &       2000 \\
 00013984003 & 2021-01-15 22:17:01 & 2021-01-16 01:41:26 &       2070 \\
 00013984004 & 2021-01-16 19:04:02 & 2021-01-16 20:54:08 &       2095 \\
 00013984005 & 2021-01-17 19:07:02 & 2021-01-17 22:26:46 &       1695 \\
 00013984006 & 2021-01-27 02:40:02 & 2021-01-27 10:43:51 &       2150 \\
 00013984010 & 2021-01-29 08:55:02 & 2021-01-30 12:19:00 &       2595 \\
 00013984011 & 2021-01-31 11:43:02 & 2021-01-31 12:09:00 &       1355 \\
 00013984012 & 2021-02-03 16:09:01 & 2021-02-03 16:39:01 &       1690 \\
 00013984014 & 2021-02-05 03:20:02 & 2021-02-06 09:50:00 &       2030 \\
 00013984015 & 2021-02-07 12:31:02 & 2021-02-07 12:58:57 &       1565 \\
 00013984016 & 2021-02-10 21:41:01 & 2021-02-10 23:29:57 &        860 \\
 00013984017 & 2021-02-11 04:03:02 & 2021-02-11 10:51:59 &       1920 \\
 00013984019 & 2021-02-15 09:57:02 & 2021-02-15 10:27:01 &       1665 \\
 00013984021 & 2021-02-18 00:00:02 & 2021-02-18 09:49:56 &       1290 \\
 00013984022 & 2021-02-20 07:48:02 & 2021-02-20 09:50:57 &       1910 \\
 00013984023 & 2021-02-22 09:11:01 & 2021-02-22 09:39:00 &       1575 \\
 00013984024 & 2021-02-24 23:14:02 & 2021-02-24 23:44:01 &       1675 \\
 00013984025 & 2021-02-26 08:43:02 & 2021-02-26 21:53:58 &       1730 \\
 00013984026 & 2021-02-28 00:31:03 & 2021-02-28 19:57:01 &       1470 \\
 00013984027 & 2021-03-02 08:17:02 & 2021-03-02 10:12:00 &       1895 \\
 00013984028 & 2021-03-05 00:08:02 & 2021-03-05 05:01:58 &       1030 \\
 00013984030 & 2021-03-26 01:32:14 & 2021-03-26 08:05:57 &       1380 \\
 00013984031 & 2021-03-28 13:54:02 & 2021-03-28 14:22:58 &       1585 \\
 00013984032 & 2021-03-30 18:22:02 & 2021-03-30 21:40:00 &        610 \\
 00013984033 & 2021-03-31 05:44:02 & 2021-03-31 06:04:00 &       1135 \\
 00013984034 & 2021-04-01 10:14:02 & 2021-04-01 10:38:58 &       1350 \\
 00013984036 & 2021-04-05 16:02:20 & 2021-04-05 22:55:01 &       2045 \\
 00013984037 & 2021-04-07 07:51:02 & 2021-04-07 22:15:00 &        730 \\
 00013984038 & 2021-04-09 01:19:02 & 2021-04-09 01:48:56 &       1610 \\
 00013984039 & 2021-04-11 20:20:27 & 2021-04-11 20:34:59 &        780 \\
 00013984040 & 2021-04-13 02:47:02 & 2021-04-13 02:55:59 &        500 \\
 00013984041 & 2021-04-15 18:16:02 & 2021-04-15 23:14:59 &       1860 \\
 00013984042 & 2021-04-17 21:15:02 & 2021-04-17 23:03:56 &       2005 \\
 00013984043 & 2021-04-21 00:00:02 & 2021-04-21 01:51:00 &       1961 \\
 00013984044 & 2021-04-23 01:19:02 & 2021-04-23 22:19:01 &       2050 \\
 00013984045 & 2021-04-25 13:42:02 & 2021-04-25 20:06:57 &       1515 \\
 00013984046 & 2021-04-29 02:18:42 & 2021-04-29 04:03:00 &        860 \\
 00013984047 & 2021-05-11 00:05:02 & 2021-05-11 09:49:57 &       1225 \\
 00013984048 & 2021-07-07 00:08:02 & 2021-07-07 17:49:58 &       2100 \\
 00013984049 & 2021-07-08 19:07:02 & 2021-07-08 23:59:59 &        390 \\
 00013984050 & 2021-07-09 04:42:02 & 2021-07-10 20:30:01 &       1775 \\
 00013984051 & 2021-07-13 21:54:02 & 2021-07-13 22:08:59 &        815 \\

\hline
\end{tabular}
\caption{\label{tab:x1}List of target of opportunity observations of \scwd~by \swift{}.}
\end{table}

We have analysed XRT and UVOT data from both \swift{} \scubed{} observations and TOO observations, in order to produce long term light-curves and pre-outburst upper limits. Observations analysis was performed utilizing the standard \swift{} HEAsoft v6.28 tools, {including \texttt{xrtpipeline} for XRT processing and extraction of spectra, and \texttt{uvotmaghist} for UVOT photometry. Extraction of XRT light-curves, include correction of detector issues such as hot pixels, hot columns and pile-up, was performed using the Python \texttt{swifttools} module\footnote{https://pypi.org/project/swifttools/}, which implements the} methods presented by \cite{Evans09}. Spectral fitting was performed utilizing XSpec \citep{Arnaud96}.

The outburst light-curve is shown in Fig.~\ref{fig:xlc}, showing a double peaked outburst, with the observed brightness peaking on 2021 January 27 {(MJD 59231.8)}, at a luminosity of $3.3 \pm 0.2 \times 10^{36}$~erg~s$^{-1}$ (0.5 - 10 keV), before fading, followed by a second period of outburst that peaked at a $1.0 \pm 0.2 \times 10^{36}$~erg~s$^{-1}$ (0.5 - 10 keV) around 2021 April 13 {(MJD 59317.1)}. From initial detection by \scubed{} to the final detection in a TOO observation taken on 2021 April 29 {(MJD 59333.1)}, the outburst is lasted approximately 120~days. Note that all luminosity measurements assume a standard SMC distance of 62.1~kpc \citep{graczyk2014}.

Following these two outburst peaks, the source faded to non-detection, and TOO observations ceased. \scubed{} monitoring of the source continued, but no detection of \scwd{} was made after the final upper TOO detection limit was measured on 2021 April 29. Combined late time observations taken in 2021 July find a $3\sigma$ upper limit of $<2 \times 10^{34}$~erg~s$^{-1}$ (0.5 - 10 keV), showing that the quiescent X-ray emission was at least 2 orders of magnitude fainter than the peak brightness. We note that \scwd{} does not appear in any X-ray catalogue prior to discovery, despite being in fields observed by both \chandra{} and \xmm{} in the past. 

\begin{figure}
	\includegraphics[width=9cm,angle=-0]{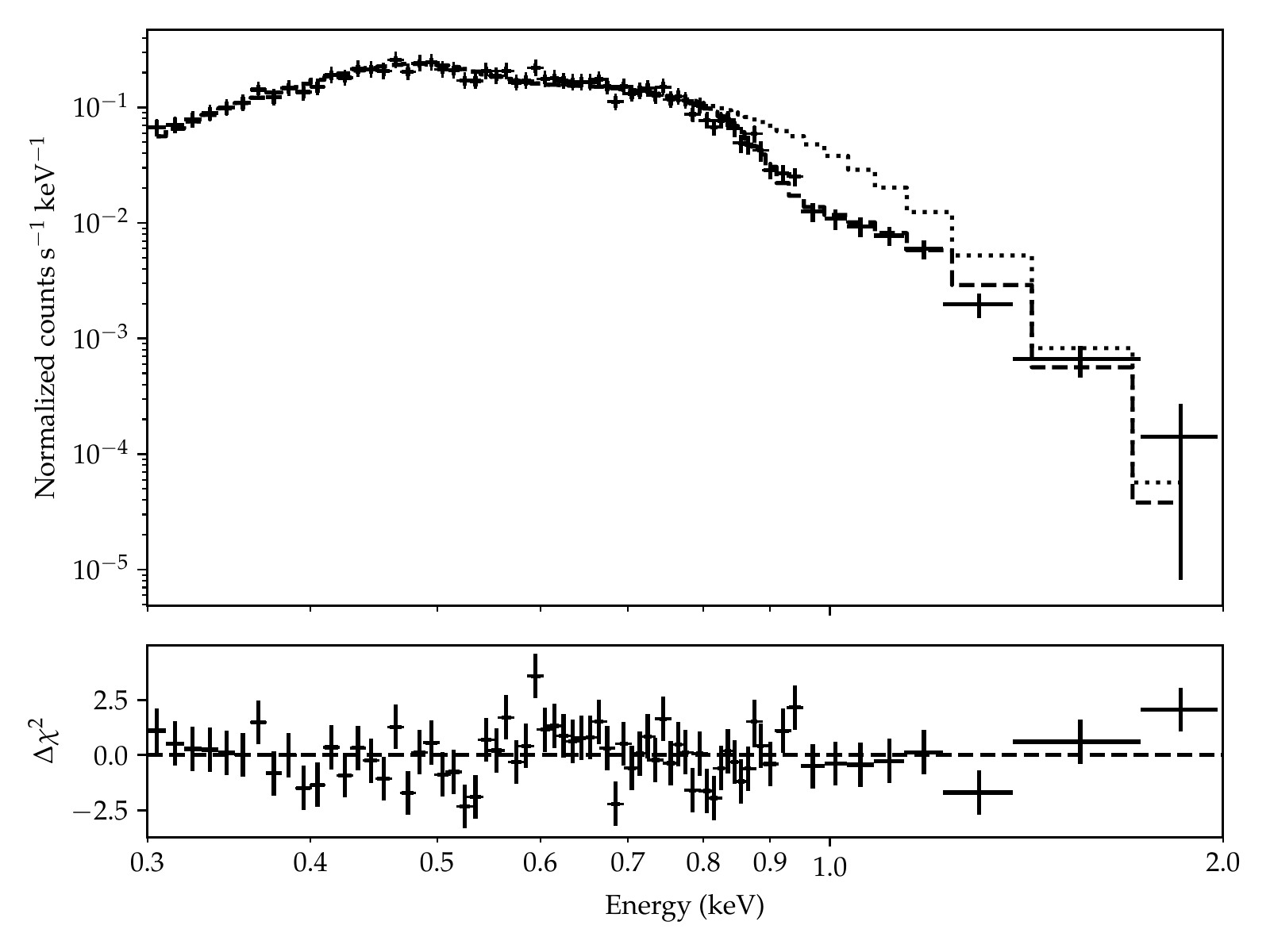}
    \caption{{Combined X-ray spectrum of \scwd{} fit with an absorbed blackbody model, with an addition of an 0.871~keV O~\textsc{viii} absorption edge.} {The dashed line shows the fitted model without the edge component.}}
    \label{fig:spectrum}
\end{figure}

In order to obtain a high quality X-ray spectrum, we combined all the \swift{} XRT TOO data. The resultant spectrum is very soft, with the majority of X-ray emission below 2 keV (see Fig.~\ref{fig:spectrum}). This is in stark contrast {to} typical \bexrb{} in which the X-ray emission is typically a hard {power-law with photon index $\simeq1$ \citep{Haberl2004}. The very soft spectrum }suggests that, analogous to the previously reported \scwdold, \scwd{} is a \bexrb{} with a WD compact object. 

The X-ray spectrum is best fit by an absorbed blackbody model {(Xspec \texttt{bbodyrad})}, with mean $N_\mathrm{H} = 0.194^{+0.026}_{-0.024} \times 10^{21}$ cm$^{-2}$ (using the Xspec \texttt{TBabs} {absorption} model with abundance set to \texttt{wilm}, e.g. \citealt{Wilms00}). This absorption is higher than the expected level of $5.9 \times 10^{20}$~cm$^{-2}$ \citep{dl1990}, however we note that \scwdold{} also showed enhanced absorption over the expected line-of-sight absorption \citep{coe2020}, which may suggest localized absorption is a common feature in BeWD binaries. \cite{Mukai17} notes that excess absorption over line-of-sight is a common feature in accreting WD binaries. 

The fit is significantly improved (reduced $\chi^2 = 0.95$ versus $2.14$) by the inclusion of an absorption edge (using the Xspec \texttt{edge} model) with a fitted edge energy {$E_\mathrm{Edge} = 0.864\pm0.011$ keV, and an edge depth $\tau_\mathrm{Edge} = 1.84 \pm 0.32$}. This is likely the signature of the O~\textsc{viii} 0.871~keV absorption edge, a common feature of WD atmospheres frequently seen in SSS (e.g. \citealt{Shimura00}).

The fitted average black body temperature is $kT_\mathrm{bb} = 96.7 \pm 4.2$~eV. Assuming spherical accretion on the WD, \cite{Mukai17} shows that for a WD with mass $1.2 \mathrm{M}_\odot$, the maximum shock temperature would be 92 eV, consistent with the measured blackbody temperature  in \scwd. Therefore the X-ray emission here is consistent with an accreting WD scenario, with the mass of the WD of $\sim1.2~\mathrm{M}_\odot$.

{In order to make an estimate of the maximum} radius of the emitting region, {we fit the spectrum of \scwd\ at peak luminosity, with the same model as above, with $N_\mathrm{H}$, $E_\mathrm{Edge}$, $\tau_\mathrm{Edge}$, and $kT_\mathrm{bb}$ all fixed at the above average values, with only normalization free. The normalization is equal to $R_\mathrm{bb}^2/D_{10}^2$ where $R_\mathrm{bb}$ is the emission radius in kilometers, and $D_{10}$ is the distance to the source in units of 10~kpc, allowing emission radius to be calculated if the distance is known. From this fit we find a blackbody emission radius of $R_{\mathrm{bb}} = 1642 \pm 83$~km, assuming the standard SMC distance of 62.1~kpc}. This value {is order-of-magnitude consistent with} the estimated $\sim4000$km radius of a $1.2~\mathrm{M}_\odot$ WD (e.g. \citealt{Pringle75}), consistent with the hypothesis that \scwd{} is a BeWD in the SMC, rather than a foreground object. 

\begin{figure}
	\includegraphics[width=9cm,angle=-0]{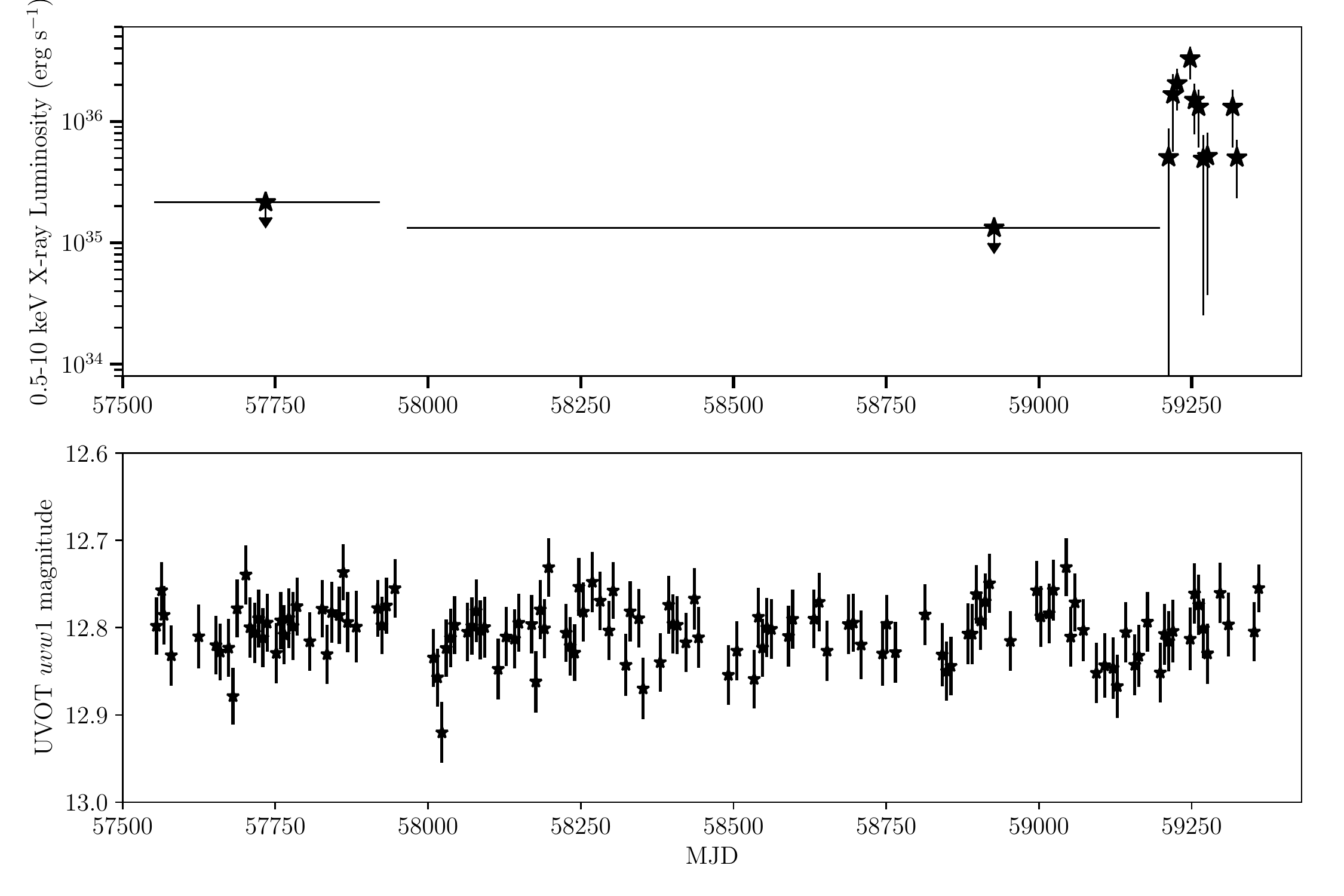}
    \caption{\scubed{} light-curve of \scwd{} showing all XRT and UVOT data. Upper limits on combined \scubed{} data taken before outburst are shown at a level of $\simeq1-2 \times 10^{35}$~erg~s$^{-1}$ (0.5 - 10 keV). UVOT {\it uvw1} light-curve from \scubed{} is statistically consistent with no variability in the UV emission from \scwd{}.}
    \label{scubed_lc}
\end{figure}

Analysis of the UVOT light-curve derived from both \scubed{} data and the TOO observations do not show any significant brightness changes in the optical/UV counterpart. Fig.~\ref{scubed_lc} shows the \scubed{} X-ray and UVOT {\it uvw1} light-curve for the entire period that \scubed{} monitoring the source, starting in 2016 June. Significantly, the outburst of \scwd{} was not preceded or accompanied by any UV brightening. Fitting a constant level model to the UVOT {\it uvw1} light-curve reveals that it is consistent with no statistically significant variability, with reduced $\chi^2 = 0.9$. Therefore there is no detectable UV signature of this outburst.

\subsection{OGLE optical photometry}

The OGLE project \citep{Udalski2015} undertakes to provide long term I-band photometry with an average cadence of 1-3 days. The star \twomass{} was observed continuously for nearly 2 decades in the I-band until COVID-19 restrictions prevented any further observations from March 2020. It is identified in the OGLE catalogue as:\\
\\
O-III: smc115.1.38 I-band\\
O-IV: smc732.06.6370 I-band\\
OIII: smc115.1.v.17 V-band\\
OIV:  smc732.06.v.7898 V-band\\

\begin{figure}
	\includegraphics[width=8cm,angle=-0]{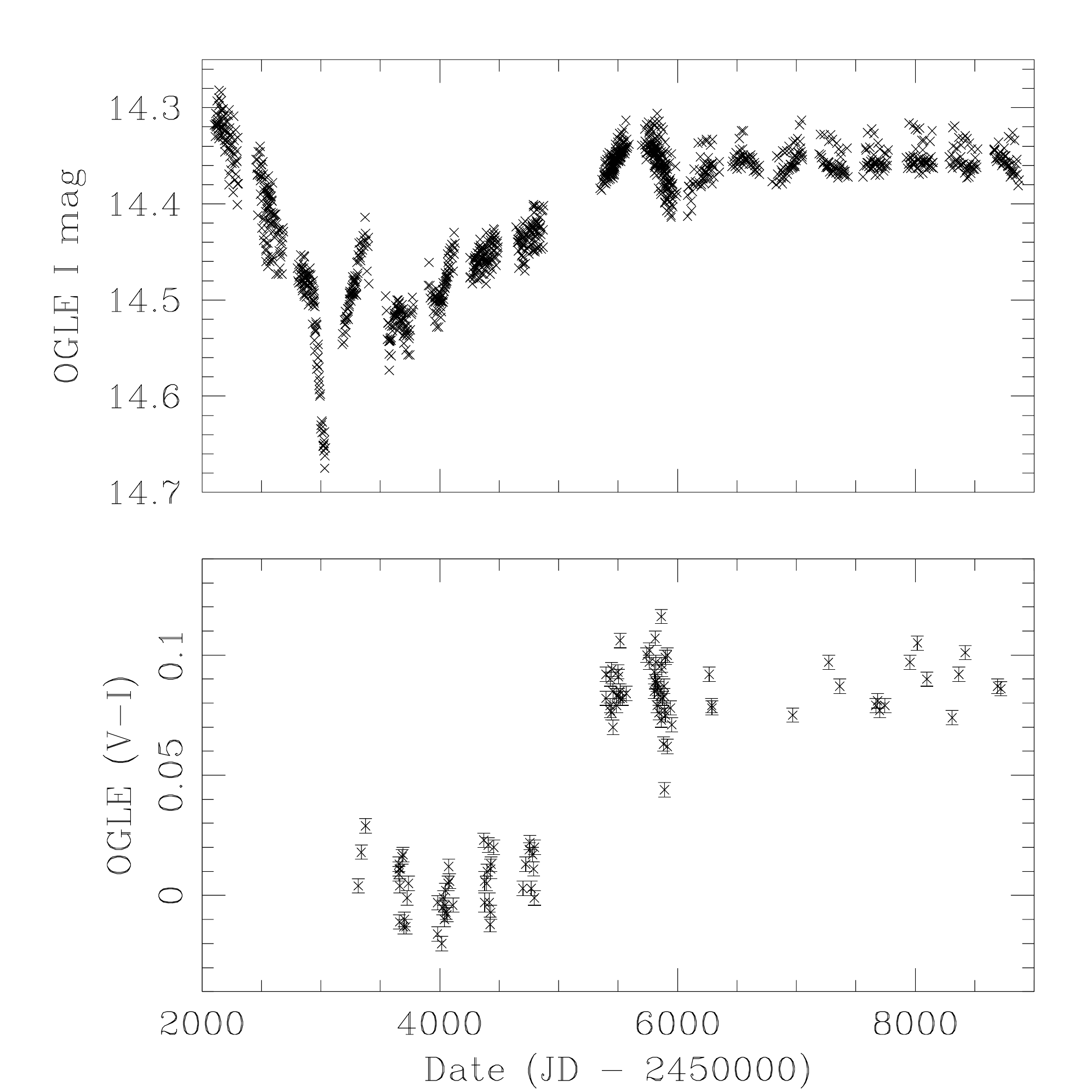}
    \caption{{Top panel : OGLE III and IV observations of the optical counterpart to \scwd. Lower panel OGLE (V-I) colour versus time.}}
    \label{fig:merged}
\end{figure}

The I band data are shown in their entirety in Fig.~\ref{fig:merged}. The overall behaviour revealed is consistent with that expected for a Be star in the SMC, showing large scale fluctuations on timescales of years. Also shown in this figure are the (V-I) colour changes observed by OGLE during this period. The source has clearly been much redder in colour since its return to its brightest state. This is discussed further in the Section~\ref{discussion} below.

\begin{figure}
	\includegraphics[width=8cm,angle=-0]{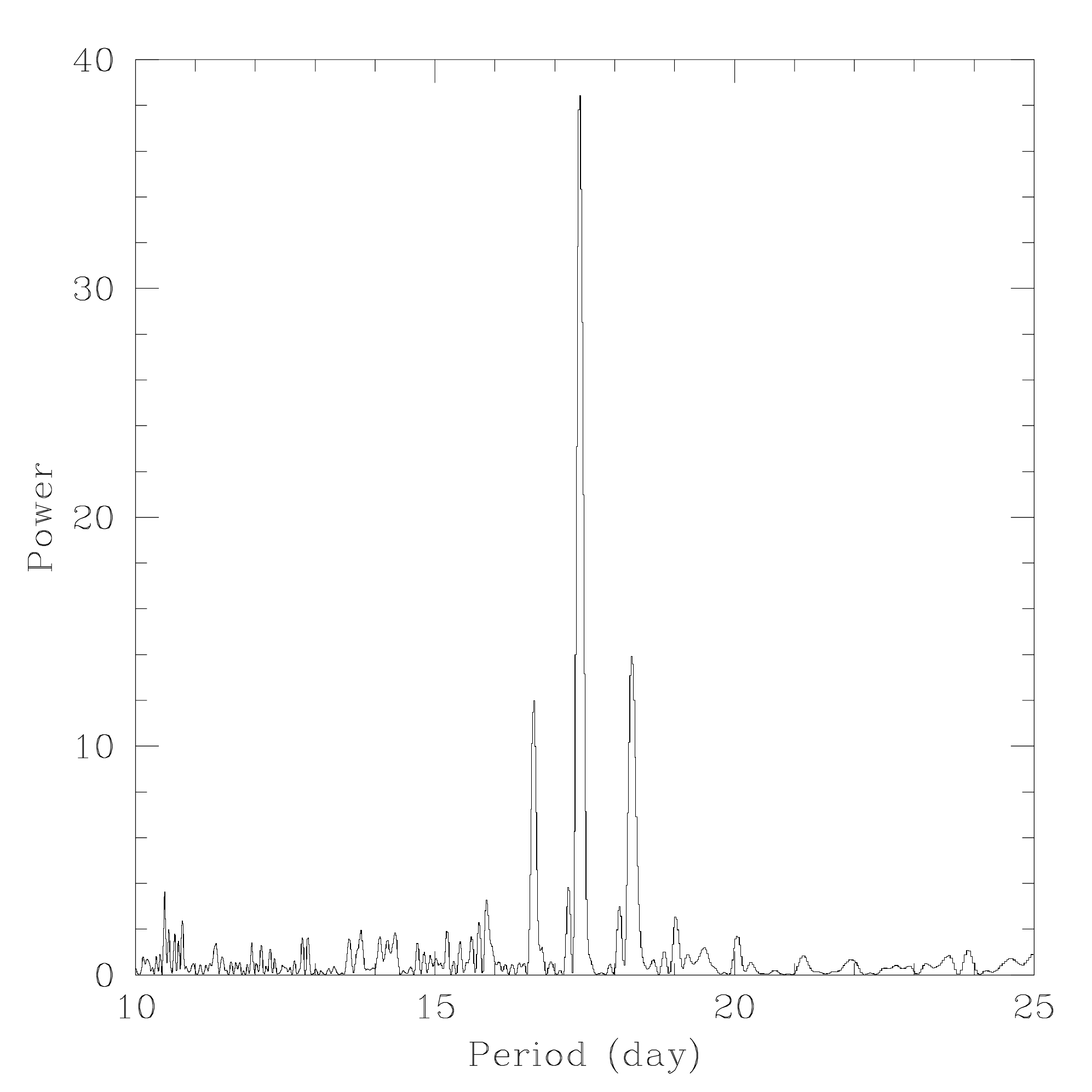}
    \caption{{Lomb-Scargle power spectrum of OGLE data of \scwd. The main peak is at 17.402~days.}}
    \label{fig:ls}
\end{figure}

Data from the most stable section of the OGLE observations, namely the 8 years after MJD 56500, were selected for a Lomb-Scargle timing analysis \citep{Lomb76,Scargle82}. The period range 2 -- 100 d was investigated and one strong peak emerged at the period of 17.402 d. The power spectrum for the range 10-25d is shown in Fig.~\ref{fig:ls}. The side peaks are the result of the true period beating with the annual sampling.

\begin{figure}
	\includegraphics[width=8cm,angle=-0]{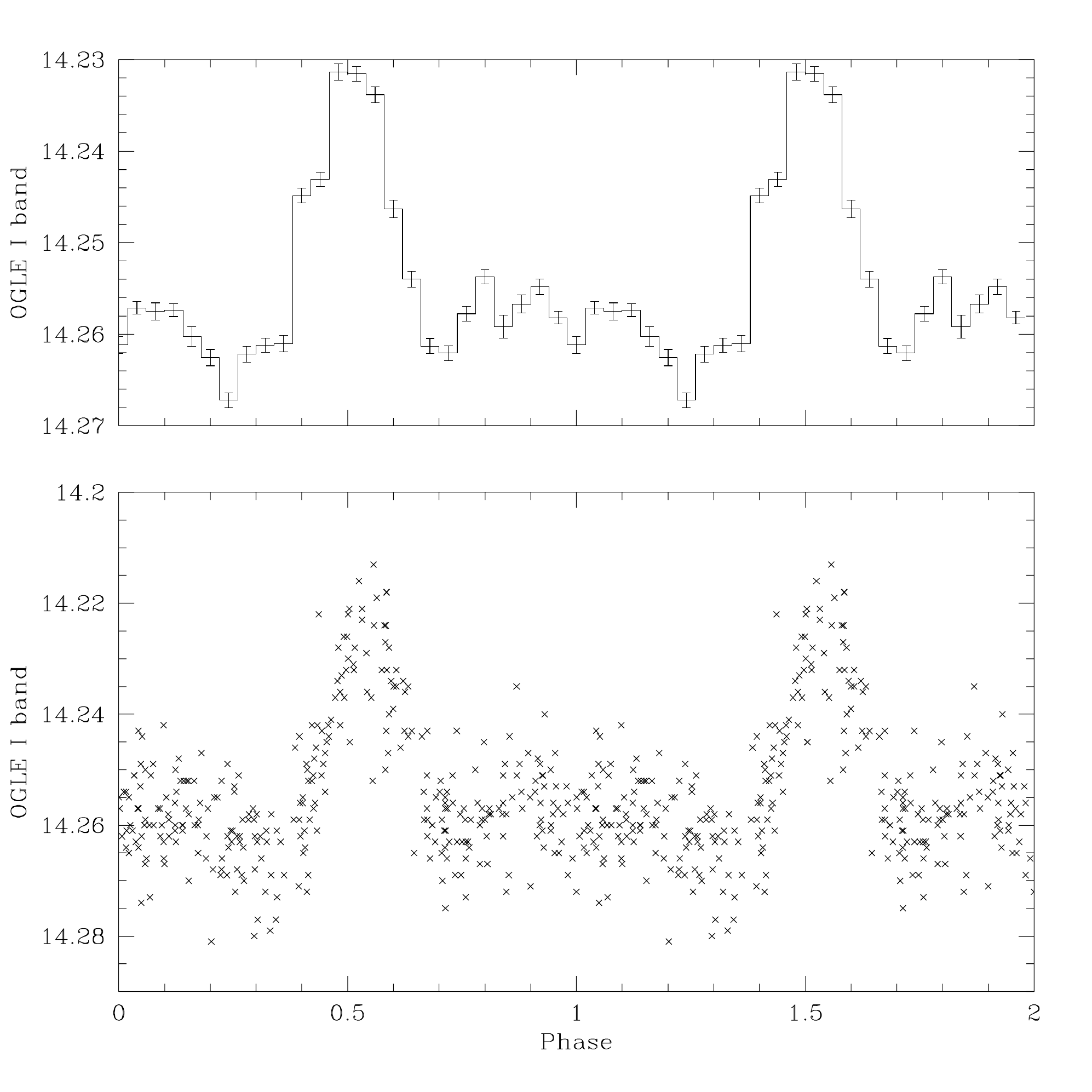}
    \caption{{OGLE I-band data folded at the period of 17.402~days. Upper panel shows the averaged values for each phase bin. Lower panel shows the scatter in the measurements.}}
    \label{fig:2panels}
\end{figure}

This subset of OGLE data was then folded at the period of 17.402 d and the resulting profile is shown in Fig.~\ref{fig:2panels}.Though there is some scatter in the individual cycles (as is shown in the lower panel of that figure) there clearly exists a dominant peak in the I-band light covering the phase range $\sim$ 0.3 of a cycle, i.e. $\sim5$~days.

The ephemeris for the time of the peak of the optical outbursts, $T_{\mathrm{opt}}$, is:

\begin{equation}
T_{\mathrm{opt}} = 56103.7 + N(17.402)~\textrm{MJD}\label{eq:1}
\end{equation}

\subsection{SALT observations}

\begin{figure}
	\includegraphics[width=8cm,angle=-0]{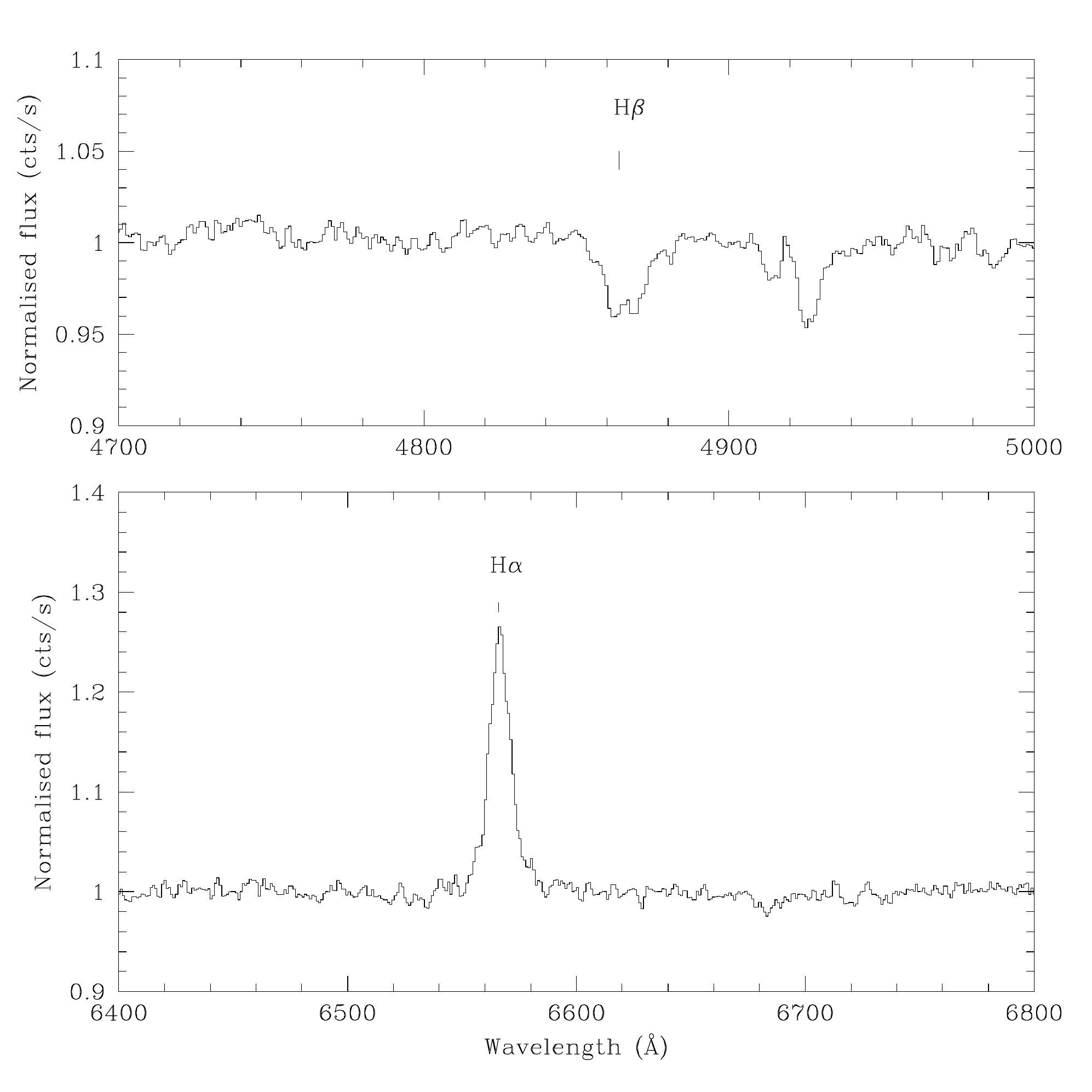}
    \caption{SALT spectra of \twomass. The upper panel shows the H$\beta$ profile, and the lower panel the H$\alpha$ profile. In both cases the predicted position of the line is indicated, adjusted for the redshift to the SMC.}
    \label{fig:salt2}
\end{figure}

\twomass{} was observed with the SALT on 2021 June 19 (JD 2459384.7) using the Robert Stobie Spectrograph \citep{Burgh03}. The exposure time was 300 s and the grating used was PG0900. The PG0900 grating produces a spectrum over the range 4000--7000 \AA, at a resolution of approximately 1700 at H$_{\alpha}$ and 1100 at H$_{\gamma}$. The SALT observations of the two main hydrogen Balmer lines are shown in Fig.~\ref{fig:salt2}. H$\alpha$ is clearly shown in emission, consistent with a single peak, and with an Equivalent Width of -3.2 $\pm$ 0.1 \AA. The H$\beta$ line is predominantly in absorption but with definite evidence of partial infilling. Such Balmer line emission is characteristic of the presence of a circumstellar disk around the B-type star.

\begin{figure*}
    \includegraphics[width=0.95\textwidth,angle=0]{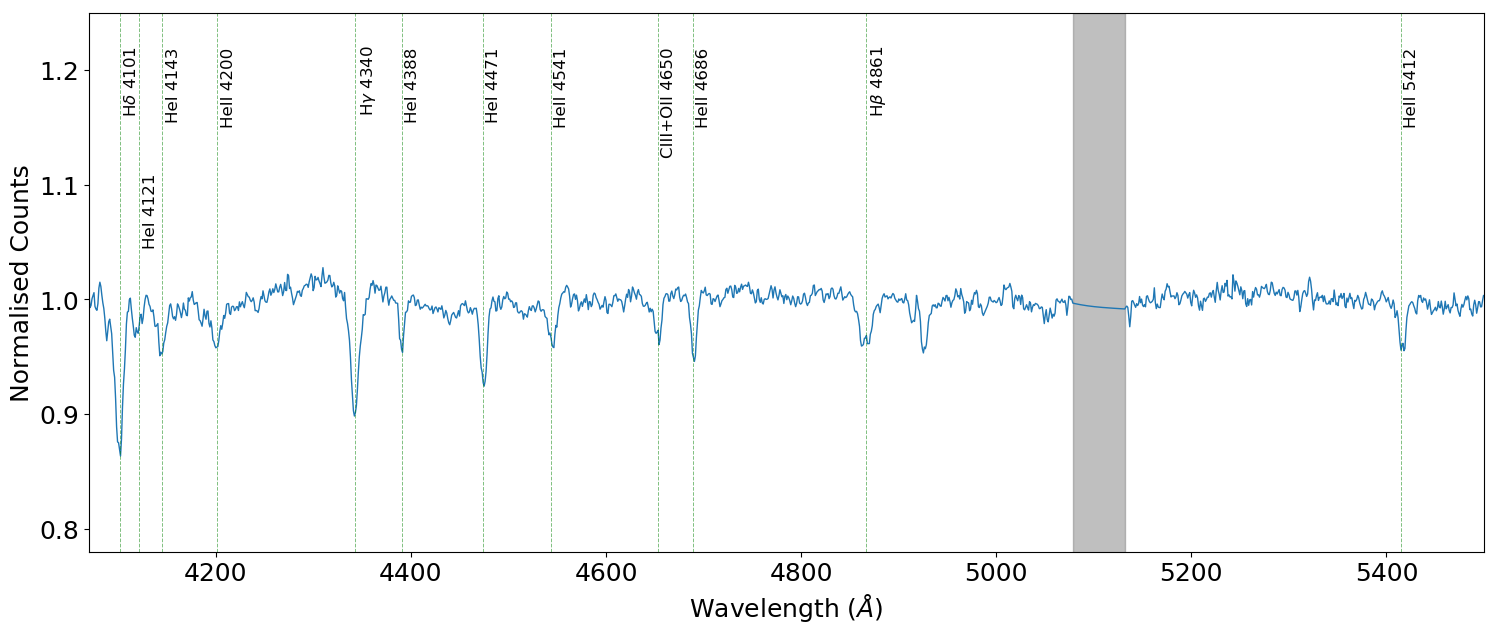}
    \caption{SALT blue end spectrum of \twomass. The grey vertical column indicates a chip gap, which has been interpolated across. {Prominent hydrogen, helium and ionized helium lines are visible, and are discussed further in the text.}}
    \label{fig:salt3}
\end{figure*}

The blue end of the spectral range is shown in Fig. \ref{fig:salt3}. The spectrum shows prominent HeII lines at 4200, 4541, 4686 and 5412 \AA, as well as {the ionized C+O blend at 4650 \AA}. These lines show that the star is a late-type O star. As outlined in \cite{wf1990} and discussed in \cite{gm2016}, the ratio of HeII 4541 / HeI 4471 is < 1, indicating a spectral type later than O7. The ratios HeII 4200 / HeI 4143 and HeII 4541 / HeI 4388 are both approximately 1, signifying a spectral type of O9. This is supported by the clear identification of CIII 4650 {(blended with the weaker OII line)} and {possible} detection of SiIV 4089 {(though we note it is very difficult to identify this line for certain above the noise of the continuum)}; both indicators of late-type O stars. It is difficult to distinguish between luminosity classes III-V due to the absence of strong metal lines in the SMC. However, our spectrum matches very well with the one presented in \cite{gm2016} ({see the bottom spectrum of their Fig. 8}). Those authors conclude an O9 IIIe classification based partly on the V-magnitude expected at the distance of the SMC. Based on our spectrum, we fully agree with their classification.

\section{Discussion}\label{discussion}

\scwd{} like many other \bexrb{} systems was discovered through X-ray outburst, in this case the outburst itself was prolonged, much longer than the measured 17.4~day orbital period, suggesting a Type-II outburst. The soft thermal spectrum and presence of an O~\textsc{viii} absorption edge, commonly seen in WD sources such as SSS, clearly identifies the compact object in this system as a WD. The proposed optical counterpart, \twomass{}, has a spectral type of O9IIIe, suggesting that \scwd{} is in fact a newly discovered BeWD binary system. \scwd{} would only be the sixth candidate BeWD identified, fourth in the SMC, following on from \scwdold{} which was also discovered by \scubed, XMMU~J010147.5-715550 \citep{sturm2012}, Suzaku J0105–72 \citep{cracco2018}, and the two Large Magellanic Cloud (LMC) BeWD candidates, XMMUJ~052016.0-692505 \citep{k2006} and RX J0527.8-6954 \citep{oliveira2010}.

\begin{figure}
	\includegraphics[width=8cm,angle=-0]{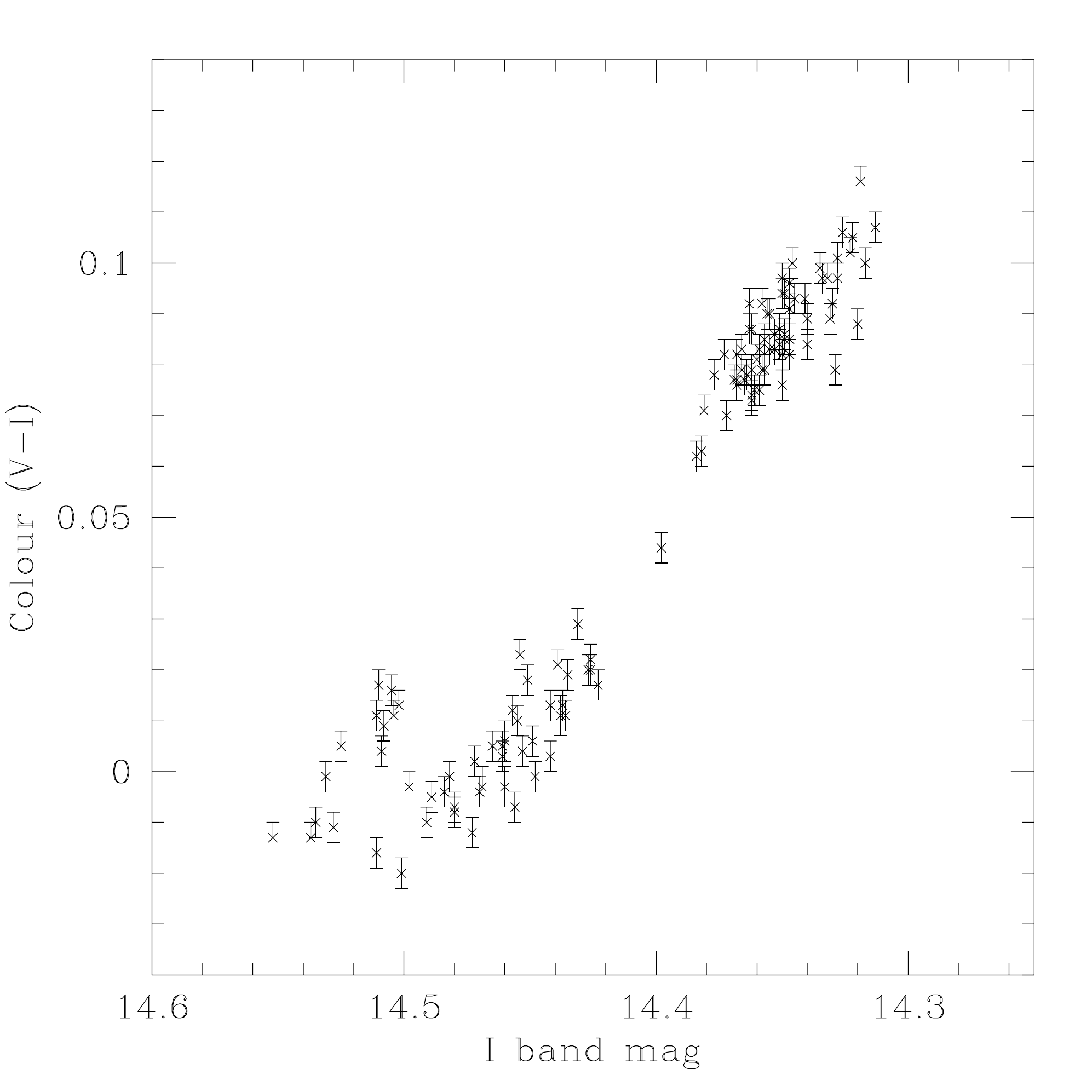}4
    \caption{{Colour-magnitude trend diagram from the OGLE data of \twomass.}}
    \label{fig:cmd}
\end{figure}

Fig.~\ref{fig:cmd} shows the strong correlation between the brightness of \twomass{} and the redness of the whole system. The observed light is arising as a result of a fixed contribution from the OB star plus a variable contribution from the,  generally cooler, circumstellar disc. Taking the spectral class of the star to be O9IIIe \citep{gm2016} with a colour of (V-I) = --0.37 \citep{pm2013}, and assuming the reddening to the SMC to be E(V-I)=0.067 \citep{skowron2021}, this implies that the underlying observed colour of the star, without any circumstellar disc, would be (V-I) $\sim$ --0.30. It is clear from Fig.~\ref{fig:merged} that the overall colour never reaches that blue extreme during the time of these OGLE observations, indicating the continual presence of a significant, but variable circumstellar disc in this system.

{\cite{Kennea2020} reported that in the case of the \scubed{} discovered \bexrb{} \scnew{}, the UVOT \textit{uvw1} brightness tracked the OGLE-\textit{I} brightness. The OGLE-\textit{I} band light-curve in Fig.~\ref{fig:2panels} taken between the June 2016 and March 2020 show that the source not showing any large scale variations during this period (excluding orbital variations). Similarly the UVOT \textit{uvw1} light-curve (Fig.~\ref{scubed_lc}) is steady during this period and after, including the period of the X-ray outburst. If \textit{uvw1} brightness is a proxy for $I$-band, then this is suggestive that the optical state remained stable after March 2020. We note that no UV enhancement is expected associate with the super-soft X-ray emission, as the extrapolated UV brightness is too faint to be detected.}

 The timings of all the XRT detections in \scwd{} were investigated, as was the strength of each detection. Folding both against the phase given in Eqn.~\ref{eq:1} does not reveal any preferences for X-ray emission at any specific binary phase. This is, perhaps, not surprising if we are seeing a classic Type II outburst. Such events are believed to occur when the circumstellar disc expands sufficiently far to encompass the whole of the neutron star orbit. Type-I outbursts, if they exist in these BeWD binary systems, are likely below the level of detection of the shallow \scubed{} exposures. 
 
\begin{figure}
	\includegraphics[width=8cm,angle=-0]{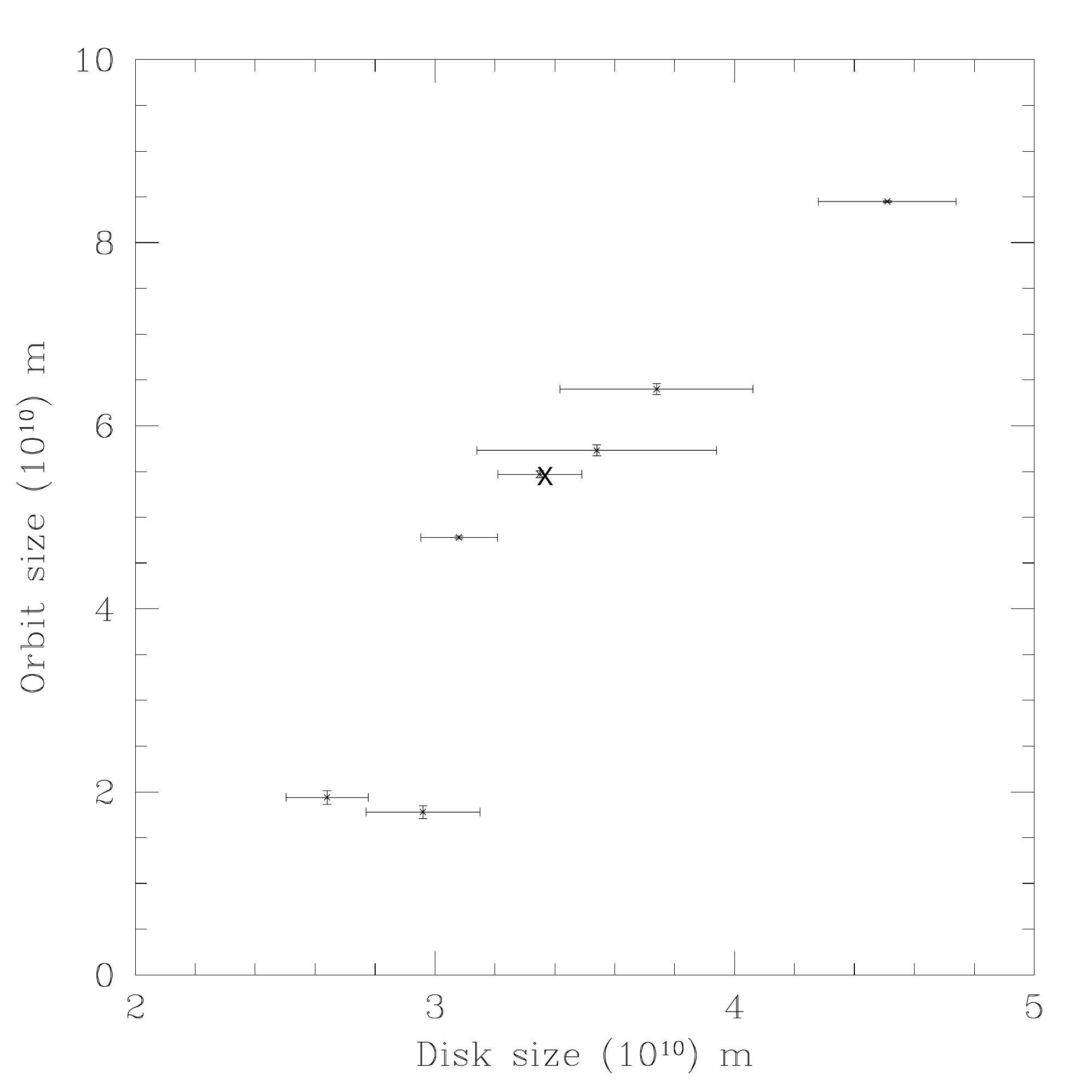}
    \caption{{The estimated circumstellar disc size versus orbital size for a sample of HMXB systems, plus that of \scwd{} (shown as a X symbol). See text for details.}}
    \label{fig:disk}
\end{figure}

 The small size of the H$\alpha$ emission seen in this system puts it amongst a group of other High Mass X-ray Binary (HMXB) systems which also show a small Equivalent Width ($\ge$ -10\AA), but still exhibit significant X-ray activity. There are five such systems known in the SMC (SXP 8.65, SXP 8.80, SMC X-1, SXP 101, SXP 2.37) and one in the Milky Way (SAX J2103.5+4545) that have well documented small H$\alpha$ EW values and known binary periods. Following the technique described in \cite{CK2015} it is possible to estimate the size of the circumstellar disks \& the size of the orbital radii, and then compare these values with that found here for \scwd. {The technique relies upon estimating the circumstellar disk size from the strength of the H$\alpha$ emission, and the size of the orbit from the orbital period and the mass of the OB star. Using that approach for \scwd ~the circumstellar disk size is estimated to be $(3.35\pm 0.14)\times 10^{10}$ m and the orbital size to be $(5.47\pm 0.17)\times 10^{10}$ m. }This result is shown in Fig.~\ref{fig:disk}. 
 
 It is immediately apparent that a strong correlation exists between these two parameters, confirming that in these tight orbital situations the orbiting compact object plays a major role in constraining the extent of the circumstellar disc. Such restrictions have been proposed by \cite{on2001} based upon their Smooth Particle Hydrodynamic simulations.
 Though the compact object in this system is proposed here to be a WD rather than a NS for the other systems, the location of \scwd{} on this diagram strongly suggests that such an object can also play a very similar role to that of the neutron star in the other systems.  However, studies of further Be white dwarf systems are needed to confirm that no significant differences actually exists.
  
A large population of BeWDs has been predicted by binary evolution modeling \citep{raguzova2001}, however despite this relatively few such systems are known. \scwdold{} became the only third such system to be discovered in the SMC \citep{coe2020}, despite the apparent overabundance of \bexrbs{} compared to the Milky Way. The fourth discovered SMC BeWD, \scwd{} shares similar properties to \scwdold, with orbital periods of $\sim20$~days, and both being discovered through apparent Type-II X-ray outbursts.

It is clear that without a sensitive, soft-band regular X-ray survey of the SMC such as \scubed, the likelihood of discovery of sources like \scwd{} and \scwdold, is much smaller. The peak fluxes of these sources, combined with their soft spectra put them below the level of detection of current all-sky survey instruments such as MAXI \citep{Matsuoka09}, Fermi/GBM \citep{Meegan09} and \swift{}/BAT \citep{Barthelmy05}. 

If Type-I outbursts are absent or too faint to be detected by \scubed{} in SMC BeWDs, to detect X-ray emission from a BeWD it needs to be undergoing a Type-II outburst, which are brighter and longer lasting than Type-I. Unfortunately Type-II outbursts are also much less frequent, with often years between outbursts. So in order to discover new BeWDs, observations need to occur in the right place, i.e. in a location where BeWDs are likely to exist and are detectable (i.e. low absorption in the line of sight), and at the right time (when they are undergoing a Type-II outburst). Given this, the relative paucity of known BeWDs can at least partially be explained due to observational bias.

This highlights the importance of sensitive soft X-ray surveys as a discovery engine of BeWD binaries. It is therefore not surprising that \scwdold{} was also detected \citep{ATel13709} by the all-sky soft X-ray survey telescope eROSITA \citep{Predehl21}. However, the focused SMC only approach of \scubed{} has proven a powerful tool for discovery of BeWD, by in 5 years doubling the known number of SMC BeWDs.

It is notable that as yet, a BeWD has not been discovered in the Milky Way, only in the SMC and LMC. This may be a selection effect due to much higher extinction rates in the plane of the Milky Way, combined with the observational biases discussed above. On the other hand the different metallicity of the Magellanic Clouds may play a crucial role in the evolution of such systems.

We note that BeWD are important systems, as they have been suggested as progenitors of peculiar overluminous Type Ia, and possibly also super luminous Type II supernovae such as SN 2006gy \cite{Ablimit21}. In addition they may also play a role as progenitors in the evolution of the proposed BH - WD systems \citep{2021sx}. In the example presented in this paper, and in the similar BeWD source studied in our previous paper \citep{coe2020}, the mass of the OB star presents a real challenge in understanding how so much mass transfer would have taken place to evolve these BeWD systems. It is clearly crucial that future observations establish an accurate estimate of their numbers in the family of evolved binary populations.

\section{Conclusions}

We report in this work the discovery, by \scubed{} of a new \bexrb{} containing a WD compact object, \scwd, which underwent a Type-II X-ray outburst over a period of 120 days. Analysis of optical and IR data from \swift, OGLE and SALT confirm the presence of a O9IIIe companion star hosting a CSD, and X-ray data are consistent with a WD at SMC distance. BeWD binary systems are a rare class of object, of which \scwd{} is only the fourth such system detected in the SMC, despite the predicted large population. We suggest that this may be due in part to the super-soft X-ray emission being only detectable using sensitive focused X-ray imaging telescopes, and only during relatively infrequent Type-II outbursts, that typically last of the order of months. This means that the likelihood of discovery of BeWD sources is greatly increased with regular sensitive soft X-ray monitoring, such as that performed by \scubed. It is likely that all-sky sensitive soft X-ray monitoring performed by eROSITA will be a significant discovery engine for BeWD binaries, perhaps discovering the first such system in the Milky Way. However, in order to increase the number of confirmed BeWD binaries, continued monitoring of the SMC by surveys like \scubed{} will prove invaluable.

\section*{Acknowledgements}

The OGLE project has received funding from the National Science Centre, Poland, grant MAESTRO 2014/14/A/ST9/00121 to AU. JAK and ZAC acknowledge support from NASA Grant NAS5-00136. PAE acknowledges UKSA support. LJT is supported by the South African National Research Foundation. Some of the observations reported in this paper were obtained with the Southern African Large Telescope (SALT) under programme 2019-2-MLT-004. The Polish participation in SALT is funded by grant no. MNiSW DIR/WK/2016/07. This work made use of data supplied by the UK \swift{} Science Data Centre at the University of Leicester.

\section*{Data Availability}

The data underlying this article will be shared on reasonable request to the corresponding author.



\bibliographystyle{mnras}
\bibliography{sc1825} 








\bsp	
\label{lastpage}
\end{document}